%% file: pret-vs-uppaal20evote.tex
\newcommand{\extended}[1]{}    
\newcommand{\short}[1]{#1}     

\documentclass{llncs}

\usepackage{amssymb,amsmath}
\usepackage{latexsym}
\usepackage{epstopdf}

\usepackage{llncsdoc}
\usepackage{framed}
\input{llncsPreamble.tex}

\usepackage{wojtek15logics,wojtek15other}

\newcommand{\Uppaal}{\textsc{Uppaal}\xspace}
\newcommand{\Pret}{Pr\^et \`a Voter\xspace}

\newcommand{\mcmas}{\text{MCMAS}\xspace}
\definecolor{lightyellow}{cmyk}{0,0,0.2,0}

\begin{document}
\pagestyle{plain} 

\title{Model Checkers Are Cool:}
\subtitle{How to Model Check Voting Protocols in \Uppaal}
\author{Wojciech Jamroga\inst{1,2} \and
Yan Kim\inst{1} \and
Damian Kurpiewski\inst{2} \and
Peter Y. A. Ryan\inst{1}}
\institute{SnT, University of Luxembourg \and
Institute of Computer Science, Polish Academy of Sciences, Warsaw, Poland}

\maketitle

\begin{abstract}
The design and implementation of an e-voting system is a challenging task.
Formal analysis can be of great help here. In particular, it can lead to a better understanding of how the voting system works, and what requirements on the system are relevant.
In this paper, we propose that the state-of-art model checker \Uppaal provides a good environment for modelling and preliminary verification of voting protocols.
To illustrate this, we present an \Uppaal model of \Pret, together with some natural extensions.
We also show how to verify a variant of receipt-freeness, despite the severe limitations of the property specification language in the model checker.
\end{abstract}

\section{Introduction}\label{sec:intro}
\input{introduction.tex}

\section{Towards Model Checking of Voting Protocols}\label{sec:towards-mcheck}
\input{preliminaries.tex}

\section{Outline of \Pret}\label{sec:pret-overview}
\input{pret-overview.tex}

\section{Modelling \Pret in \Uppaal}\label{sec:pret-model}
\input{model.tex} 

\section{Verification}\label{sec:verification}
\input{verification.tex}

\section{Replicating Pfitzmann's Attack}\label{sec:model-extensions}
\input{extensions.tex}

\section{Related Work}\label{sec:related}
\input{relatedwork.tex}

\section{Conclusions}\label{sec:conclusions}
\input{conclusions.tex}

\bibliographystyle{abbrv}
\bibliography{wojtek,wojtek-own}

\end{document}

%% file: llncsPreamble.tex
\usepackage[utf8]{inputenc}
\usepackage[T1]{fontenc}

\usepackage{amsfonts,amssymb,amsmath}
\usepackage{graphicx}
\usepackage{subfigure}
\usepackage{textcomp}
\usepackage[usenames]{color}
\usepackage[mathscr]{eucal}
\usepackage{multirow}
\usepackage{algorithm}
\usepackage{algorithmic}
\usepackage{paralist}
\usepackage[english]{babel}
\usepackage{threeparttable}
\usepackage{booktabs}
\usepackage{latexsym}
\usepackage{caption}
\usepackage{tikz}
\usetikzlibrary{arrows,automata,shapes,snakes,backgrounds,petri,mindmap,fit,positioning,decorations.pathmorphing}
\usepackage{pgfplots}
\usepgfplotslibrary{groupplots}
\usepackage{varwidth}

\usepackage{xspace}
\usepackage{url}

\newcommand{\zgroup}[1][p]{ \mathbb{Z}^{*}_{#1}}

{\begin{figure*}[#1]
\centering
\begin{minipage}{#2}
\begin{algorithm}[H]}%
{\end{algorithm}
\end{minipage}
\end{figure*}}

\newcommand{\CASE}[1]{\STATE \textbf{case} #1\textbf{:} \begin{ALC@g}}
\newcommand{\ENDCASE}{\end{ALC@g}}

\newcommand{\DEFAULT}{\STATE \textbf{default:} \begin{ALC@g}}
\newcommand{\ENDDEFAULT}{\end{ALC@g}}
\newcommand{\DEFAULTLINE}[1]{\STATE \textbf{default:} }

\definecolor{mygreen}{rgb}{0,0.6,0}
\definecolor{mygray}{rgb}{0.5,0.5,0.5}
\definecolor{mymauve}{rgb}{0.58,0,0.82}
\usepackage{listings}

%% file: introduction.tex

The design and implementation of a good e-voting system is highly challenging. Real-life systems are notoriously complex and difficult to analyze. Moreover, elections are \emph{social} processes: they are run by humans, with humans, and for humans, which makes them unpredictable and hard to model. Last but not least, it is not always clear what \emph{good} means for a voting system. A multitude of properties have been proposed by the communities of social choice theory (such as Pareto optimality and nonmanipulability), as well as secure voting (ballot secrecy, coercion-resistance, voter-verifiability, etc.). The former are typically set for a very abstract view of the voting procedure, and consequently miss many real-life concerns. For the latter ones, it is often difficult to translate the informal intuition to a formal definition that is commonly accepted.

In a word, we deal with processes that are hard to understand and predict, and evaluate them against criteria that are not exactly clear.
Formal analysis can be of great help here: perhaps not in the sense of providing the ultimate answers, but rather to strengthen our understanding of both how the voting system works and how it should work.
The main goal of this paper is to propose that model checkers from distributed and multi-agent systems can be invaluable tools for such an analysis.

\para{Model checkers and \Uppaal.}
Much research on model checking focuses on the design of logical systems for a particular class of properties, establishing their theoretical characteristics, and development of verification algorithms.
This obscures the fact that a model checking framework is valuable as long as it is actually \emph{used} to analyze something.
The analysis does not necessarily have to result in a ``correctness certificate''\extended{ of the system under scrutiny}.
A readable model of the system, and an understandable formula capturing the requirement are already of substantial value.

In this context, two features of a model checker are essential. On the one hand, it should provide a \emph{flexible model specification language} that allows for modular and succinct specification of processes. On the other hand, it must offer a \emph{good graphical user interface}. Paradoxically, tools satisfying both criteria are rather scarce.
Here, we suggest that the state of the art model checker \Uppaal can provide a nice environment for modelling and preliminary verification of voting protocols and their social context. 
To this end, we show how to use \Uppaal to model a voting protocol of choice (in our case, \Pret), and to verify some requirements written in the temporal logic \CTL.

\para{Contribution.}
The main contribution of this paper is methodological: we demonstrate that specification frameworks and tools from distributed and multi-agent systems can be useful in analysis and validation of voting procedures.
An additional, technical contribution consists in a reduction from model checking of temporal-epistemic specifications to purely temporal ones, in order to verify a variant of receipt-freeness despite the limitations of \Uppaal.

\para{Structure of the paper.}
We begin by introducing the main idea behind modelling and model checking of multi-agent systems, including a brief introduction to \Uppaal (Section~\ref{sec:towards-mcheck}).
In Section~\ref{sec:pret-overview}, we provide an overview of \Pret, the voting protocol that we will use for our study.
Section~\ref{sec:pret-model} presents a multi-agent model of the protocol; some interesting extensions of the model are proposed in Section~\ref{sec:model-extensions}.
We show how to specify simple requirements on the voting system, and discuss the output of model checking in Section~\ref{sec:verification}. The section also presents our main technical contribution, namely the model checking reduction that recasts knowledge-related statements as temporal properties.
We discuss related work in Section~\ref{sec:related}, and conclude in Section~\ref{sec:conclusions}.

%% file: preliminaries.tex

Model checking is the decision problem that takes a model of the system and a formula specifying correctness, and determines whether the model satisfies the formula.
This allows for a natural separation of concerns: the model specifies how the system is, while the formula specifies how it should be.
Moreover, most model checking approaches encourage systematic specification of requirements, especially for the requirements written in modal and temporal logic.
In that case, the behavior of the system is represented by a transition network, possibly with additional modal relations to capture e.g. the uncertainty of agents. The structure of the network is typically given by a higher-level representation, e.g., a set of agent templates together with a synchronization mechanism.

We begin with a brief overview of \Uppaal, the model checker that we will use in later sections. A more detailed introduction can be found in~\cite{Behrmann04uppaal-tutorial}.

\subsection{Modeling in \Uppaal}\label{sec:uppaal-modeling}

The system model in \Uppaal consists of a set of concurrent processes.
The processes are defined by templates, each possibly having a set of parameters.
The parameterized templates are used for defining a large number of almost identical processes.
Every template consists of \emph{locations}, \emph{edges}, and optional local declarations.
An example template is shown in Figure~\ref{fig:voter}; we will use it to model the behavior of a voter.

Locations are graphically represented as circles.
\textit{Initial} locations are marked by a double circle.
\textit{Committed} locations are marked by the circled letter C.
If any process is in a committed location, then the next
transition must involve an edge from one of the committed
locations.
Those are used for creating atomic sequences or
encoding synchronization between more than two components.

The edges are annotated by selections (in yellow), guards (green), synchronizations (teal), and updates (blue).
The syntax of expressions mostly coincides with that of C/C++.
\textit{Selections} bind the identifier to a value from the given range in a nondeterministic way.
\textit{Guards} enable the transition if and only if the guard condition evaluates to true.
\textit{Synchronizations} allow processes to synchronize over a common channel \texttt{ch} (labeled \texttt{ch?} in the receiver process and \texttt{ch!} for the sender).
Note that a transition on the side of the sender can be fired only if there exists an enabled transition on the receiving side labeled with the same channel identifier, and vice versa.
\textit{Update} expressions are evaluated when the transition is taken.
For a synchronizing transition, the update expressions on the sender side are executed before the receiver ones.
Straightforward value passing over a channel is not allowed; instead, one has to use shared global variables for the transmission.
\extended{In between writing and reading update expressions, a committed location will be used.}

For convenience, we will place the selections and guards at the top or left of an edge,
and the synchronizations and updates at the bottom/right.

%



\subsection{Specification of Requirements}\label{sec:uppaal-formulas}


To specify requirements, \Uppaal uses a fragment of the temporal logic \CTL~\cite{Emerson90temporal}. \CTL allows for reasoning about the possible execution paths of the system by means of the \emph{path quantifiers} $\Epath$ (``there is a path'') and $\Apath$ (``for every path''). A path is a maximal\footnote{
  I.e., infinite or ending in a state with no outgoing transitions. }
sequence of states and transitions.
To address the temporal pattern on a path, one can use the \emph{temporal operators} $\Next$ (``in the next moment''), $\Always$ (``always from now on''), $\Sometm$ (``now or sometime in the future''), and $\Until$ (``until'').
For example, the formula
$\Apath\Always\big(\prop{has\_ballot_i} \then \Apath\Sometm(\prop{voted_{i,1}} \lor \dots \lor \prop{voted_{i,k}})\big)$ expresses that, on all paths, whenever voter $i$ gets her ballot form, she will eventually cast her vote for one of the candidates $1,\dots,k$.
Another formula, $\Apath\Always\neg\prop{punished_i}$ says that voter $i$ will never be punished by the coercer.

More advanced properties usually require a combination of temporal modalities with \emph{knowledge operators $K_a$}, where $K_a\phi$ expresses ``agent $a$ knows that $\phi$ holds.''
For example, formula $\Epath\Sometm(\prop{results} \land \neg\prop{voted_{i,j}} \land \neg K_c \neg\prop{voted_{i,j}})$ says that the coercer $c$ might not know that voter $i$ hasn't voted for candidate $j$, even if the results are already published. 
Moreover, $\Apath\Always(\prop{results} \then \neg K_c \neg\prop{voted_{i,j}})$ expresses that, when the results are out, the coercer won't know that the voter refused to vote for $j$.
Intuitively, both formulas capture different strength of receipt-freeness for a voter who has been instructed to vote for candidate $j$.

%% file: pret-overview.tex

In this paper, we use \Uppaal for modeling and analysis of a voting protocol.
The protocol of choice is \Pret. A short overview of \Pret is presented here;
the full details can be found, for example, in~\cite{Ryan10atemyvote\extended{,Feng17evoting}}.

Most verifiable voting systems follow the pattern: at the time of casting, an encryption or encoding of the vote is created and posted to a public bulletin board. The voter can later check that her encrypted ballot appears correctly. The set of posted, ballots are then processed in some verifiable way to reveal the tally or outcome. Much of this is effectively a secure distributed computation, and as such is well-established and understood cryptography. The really challenging bit, because it involves interactions between the users and the system, is the creation of the encrypted ballots. This has to be in a way that assures the voter that her vote is correctly embedded, while avoiding any coercion or vote buying threats.

The key innovation of the \Pret approach is to encode the vote using a randomised candidate list. This contrasts with earlier verifiable schemes that involved the voter inputting her selection to a device that then produces an encryption of the selection. Here what is encrypted is the candidate order which can be generated and committed in advance, and the voter simply marks her choice on the ballot in the traditional manner.

\begin{figure}[t]
\begin{center}
\begin{tabular}{c@{\qquad}c@{\qquad\qquad\qquad\qquad}c@{\qquad}c}
  \begin{tabular}{c}
  (a)
  \end{tabular}
&
  \begin{tabular}{|l|p{36pt}|} \hline
  Obelix & \\ \hline Idefix    & \\ \hline Asterix & \\ \hline
  Panoramix &
  \\ \hline
       & {$7304944$} \\ \hline
  \end{tabular}
&
  \begin{tabular}{c}
  (b)
  \end{tabular}
&
  \begin{tabular}{|c|} \hline
   \\ \hline X
   \\ \hline
   \\ \hline
   \\ \hline ${7304944}$ \\ \hline
  \end{tabular}
\end{tabular}
\end{center}
\caption{(a) \Pret ballot form;\ (b) Receipt encoding a vote for
``Idefix''} \label{fig:pret}
\end{figure}

Suppose that our voter is called Anne. At the polling station, Anne registers and chooses at random a
ballot form sealed in an envelope and saunters over to the booth. An example of such a form is
shown in Figure~\ref{fig:pret}a.
In the booth, Anne extracts her ballot form from the envelope and
makes her selection in the usual way by placing a cross in the right
hand column against the candidate of her choice (for approval or ranked voting, she marks her selection or
ranking against the candidates). Once her selection has been made,
she separates the left and right hand strips\extended{ along a thoughtfully
provided perforation} and discards the left hand strip. She is left
with the right hand strip which now constitutes her \emph{privacy
protected receipt}, as shown in Figure~\ref{fig:pret}b.

Anne now exits the booth clutching her receipt, returns to the registration desk, and casts her receipt. Her receipt is placed over an
optical reader or similar device that records the random value at
the bottom of the strip and records in which cell her X is marked.
Her original, paper receipt is digitally signed and franked and
returned to her to keep and later check that her vote is correctly recorded on the bulletin board.
The randomisation of the candidate list on each ballot form ensures
that the receipt does not reveal the way she voted, so ensuring the
secrecy of her vote. Incidentally, it also removes any bias towards
the candidate at the top of the list that can occur with a fixed
ordering.

The value printed on the bottom of the receipt is what enables extraction of the vote during the
tabulation phase: buried cryptographically in this value is the
information needed to reconstruct the candidate order and so extract
the vote encoded on the receipt. This information is encrypted with
secret keys shared across a number of tellers. Thus, only a
threshold set of tellers acting together are able to interpret the
vote encoded in the receipt. In practice, the value on the receipt will be the output of an agreed cryptographic hash of the ciphertext committed to the bulletin board.

After the election, voters\extended{ (or perhaps proxies acting on their
behalf)} can visit the secure Web Bulletin Board (WBB) and confirm
that their receipts appear correctly. Once any discrepancies are
resolved, the tellers take over and perform anonymising mixes and
decryption of the receipts. At the end, the plaintext votes will be posted in secret shuffled order, or in the case of homomorphic tabulation, the final result is posted. All the processing of the votes can be made universally verifiable, i.e., any observer can check that no votes were manipulated. These are carefully designed so as not to impinge on
ballot privacy.

\Pret brings several advantages in terms of privacy and dispute resolution. Firstly, avoidance of side channel leakage of the vote from the encryption device. Secondly, improved dispute resolution: ballot assurance is based on random audits of the ballot forms, which can be performed by the voter or independent observers. A ballot form is either well-formed, i.e. the plaintext order matches the encrypted order, or not. This is independent of the voter or her choice, hence there can be no dispute as to what choice the voter provided. Such disputes can arise in Benaloh challenges and similar cut-and-choose style audits. Furthermore, auditing ballots does not impinge on ballot privacy, as nothing about the voter or the vote can be revealed at this point.

%

%% file: model.tex

In this section, we present how the components and participants of \Pret can be modelled in \Uppaal.
To this end, we give a\extended{ detailed} description of each module template, its elements, and their interactions.
The templates represent the behavior of the following types of agents: \emph{voters}, \emph{coercers}, \emph{mix tellers}, \emph{decryption tellers}, \emph{auditors}, and the \emph{voting infrastructure}.
For more than one module of a given type, an identifier $i=0,1,\dots$ will be associated with each instance.

The code of the model is available at \url{https://github.com/pretvsuppaal/model}.
\short{
  Here, we present in detail only the Voter template. The details of the other modules can be found in the extended version of the paper, available at \url{https://github.com/pretvsuppaal/extended}.
}

\extended{
  In the model, we use the \textit{ElGamal} encryption algorithm, which allows for re-encryption.
  The public key is a tuple $(p, \alpha, \beta)$, where $\alpha$ is a
  generator of group $\zgroup$, $\beta = \alpha^{k}$
  and $k$ is a secret (private key).
  A plain-text $m$ encrypted with public key $PK$
  with randomness $y$ will be noted as $E_{PK}(m,y)$,
  if the value of $y$ is not known,
  then it will be noted as $E_{PK}(m,*)$ ( or simply $E_{PK}(m)$ ) instead.
  A ciphertext $c$ decrypted with private key $K$ is noted by $D_K(c)$.

  \subsection{Environment}\label{sec:global-shared-scope}

  We begin by an overview of the shared environment of action, i.e., the data structures and variables shared by the modules.
  To capture a possibly repeated block of atomic update expressions,
  some procedures are introduced.
  This will allow for more complex expressions and
  a shorter, reader-friendly form of labels.

  The environment includes some global read-only configuration variables:
  \begin{itemize}
  \item \texttt{c\_total[=3]}: the number of candidates,
  \item \texttt{v\_total[=3]}: the number of voters,
  \item \texttt{mt\_total[=3]}: the number of mix tellers,
  \item \texttt{dt\_total[=3]}: the number of decryption tellers,
  \item \texttt{dt\_min[=2]}: the number of participants needed to reconstruct a secret key,
  \item \texttt{z\_order[=7]}: an order of cyclic group $\zgroup$,
  \item \texttt{pk[=(3,6)]}: the pair of generator $\alpha$ of group $\zgroup$ and $\beta=\alpha^k\ (mod\ p)$, where $k$ is a secret key.
  \end{itemize}

  From the model configuration values, we derive some auxiliary variables, such as
  lists of permutations of the batch terms \texttt{P\_b},
  list of permutations of the candidates \texttt{P\_c},
  list of cyclic shifts of the candidates \texttt{S\_c},
  lookup table \texttt{dlog}, that maps onion to its seed
  ($\pm$ possible candidate choice),
  list of combinations to select a subset of the batch intended for audit \texttt{audit\_ch},
  and list of ways of splitting that in two (odd and even)
  \texttt{audit\_lr}.

}
To facilitate readability and manageability of the model code, we define some data structures and type name aliases based on the configuration variables:
\begin{itemize}
\item \textbf{Ciphertext}: a pair $(y_1,y_2)$, representing some cipher-text.
\item \textbf{Ballot}: a pair $(\theta,cl)$ of onion $\theta=E_{PK}(s,*)$
and candidate list $cl=\pi(s)$, where $s$ is a seed associated with
the ballot, $\pi:\ \mathbb{R}\rightarrow Perm_C$
is a function that associates a seed with a permutation of the candidates.
To implement basic absorption of marked cell index into the onion,
we will only use cyclic shifts of base candidate order and
use a Voter's id as a seed of associated with her ballot.

\item \textbf{Receipt}: a pair $(\theta,r)$ of onion $\theta$ and
an index $r$ of marked cell. It can be used to verify if
a term was recorded and if it was done correctly.

\item  \textbf{c\_t}: an integer with range [0, c\_total), a candidate;
\item  \textbf{v\_t}: an integer with range [0, v\_total), a voter;
\item  \textbf{z\_t}: an integer with range [0, z\_total), an element of
$\zgroup$.
\end{itemize}

\extended{
  \smallskip\noindent
  The writable global variables include:
  \begin{itemize}
  \item \texttt{board}: a 2-dimensional list of ciphertexts, representing the web bulletin board.
  The first column is reserved for the batch of onions with absorbed indices from the receipt. The next $(mt\_total\cdot2)$-columns store re-encryption mixes. The remaining $(dt\_min-1)$-columns store the intermediate results of threshold decryption; the last one holds the decrypted message.

  \item \texttt{mixes}: encodes which mix teller has a turn at the moment;
  \item \texttt{dt\_curr}: encodes the number of currently participating decryption tellers;
  \item \texttt{decryptions}: the number of decryptions made.
  \end{itemize}
  \extended{
    There is also a set of globally shared variables used to simulate the passing of a value through a channel; their use will be specified in transition descriptions.
  }

  \smallskip\noindent
  Finally, the global procedures are:
  \begin{itemize}
  \item \texttt{zpow(a,b)}: returns an element $a^b$ in $\zgroup$;
  \item \texttt{encr(m,r)}: returns a ciphertext $E_{PK}(m,r)$;
  \item \texttt{decr(c,k)}: returns a message $D_k(c)$.
  \end{itemize}
}

\begin{figure}[t]
	\hspace{-.7cm}\includegraphics[width=1.1\textwidth]{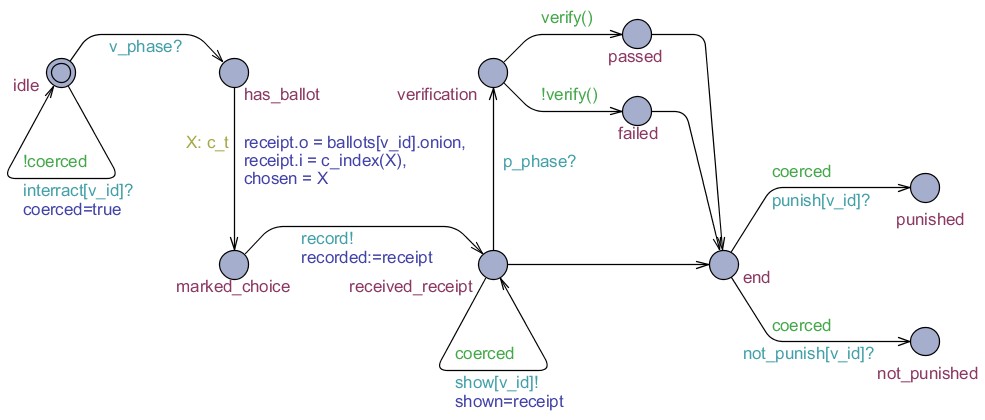}
	\caption{Voter template for the model of \Pret}\label{fig:voter}		
\end{figure}

\subsection{Voter Template}\label{sec:voter}

The structure of the Voter template is shown in Figure~\ref{fig:voter}.
The idea is that while the voter waits for the start of election she might be subject to coercion.
When the ballots are ready, the voter selects a candidate, and transmits the receipt to the system.
Then she decides if she wants to check how her vote has been recorded, and if she wants to show the receipt to the coercer.
If coerced, she also waits for the Coercer's decision to punish her or refrain from punishment.

\smallskip\noindent
The module includes the following private variables:
\begin{itemize}
\item
  \texttt{receipt}: instance of \texttt{Receipt}, obtained after casting a vote;
\item
  \texttt{coerced[=false]}: a Boolean, indicating if coercer has established a
  contact;
\item
  \texttt{chosen} - integer value of chosen candidate.
\end{itemize}

\smallskip\noindent
Moreover, the following procedures are included:
\begin{itemize}
\item
  \texttt{c\_index(target)}: returns an index, at which \texttt{target}
  can be found on the candidate list of a ballot;
\item
  \texttt{verify()}: returns $true$ if voter's \texttt{receipt} can be found
  on the board list, else returns $false$.
\end{itemize}

\smallskip\noindent
States:
\begin{itemize}
\item
  \emph{idle}: waiting for the election, might get contacted by coercer;
\item
  \emph{has\_ballot}: obtained the ballot form;
\item
  \emph{marked\_choice}: marked an index of chosen candidate (and destroyed
  left hand side with candidate list);
\item
  \emph{received\_receipt}: receipt is obtained and might be shown to the coercer;
\item
  \emph{verification}: decided to verify the receipt;
\item
  \emph{passed}: has a confirmation that the receipt appears correctly;
\item
  \emph{failed}: has an evidence that the receipt either does not
  appear or appears incorrectly (in case of index absorption both
  cases are the same);
\item
  \emph{end}: the end of the voting ceremony;
\item
  \emph{punished}: has been punished by the Coercer;
\item
  \emph{not\_punished}: has not been punished by the Coercer.
\end{itemize}

\begin{figure}[t]	
	\centering
	\includegraphics[width=0.5\textwidth]{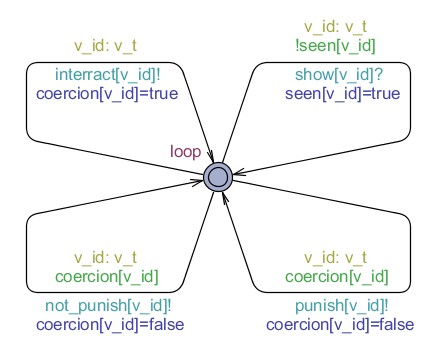}
	\caption{Coercer template}\label{fig:coercer}		
\end{figure}

\smallskip\noindent
Transitions:
\begin{itemize}
\item
  \emph{idle$\rightarrow$idle}: if was not already coerced, enable transition;
  if taken, then set \texttt{coercion} to $true$;
\item
  \emph{idle$\rightarrow$has\_ballot}: enabled transition;
  if taken, then voter acquired a ballot form;
\item
  \emph{has\_ballot$\rightarrow$marked\_choice}: mark a cell with
  chosen candidate;
\item
  \emph{marked\_choice$\rightarrow$received\_receipt}:
  send \texttt{receipt} to the System
  over channel \texttt{record} using
  shared variable \texttt{recorded};
\item
  \emph{received\_receipt$\rightarrow$received\_receipt}:
  if was coerced, enable transition;
  if taken, then pass the \texttt{receipt} to Coercer
  using shared variable \texttt{shown};
\item
  \emph{received\_receipt$\rightarrow$verification}: enabled transition;
  if taken, then Voter decided to verify whether receipt appears on board;
\item
  \emph{(received\_receipt {\normalfont ||} passed {\normalfont ||} failed)$\rightarrow$end}:
  end of voting ceremony for voter;
\item
  \emph{end$\rightarrow$punished}: if was coerced, enable transition;
  if taken, then Voter was punished by Coercer;
\item
  \emph{end$\rightarrow$not\_punished}: if was coerced, enable transition;
 if taken, then Voter was not punished by Coercer.
\end{itemize}

\subsection{Coercer}\label{sec:coercer}

The coercer can be thought of as a party that can influence voters by forcing them to obey certain instructions.
To enforce this, the Coercer can punish the voter\extended{, or refrain from the punishment}.
The structure of the Coercer is presented in Figure~\ref{fig:coercer}\short{; see the extended version of the paper at \url{https://github.com/pretvsuppaal/extended} for the technical details}.

\extended{
    \smallskip\noindent
    Private variables:
    \begin{itemize}
    \item \texttt{coercion}: a Boolean list used to keep track of the voters that have been coerced;
    \item \texttt{seen}: a Boolean list, that indicates if the voter has shown a proof of her vote.
    \end{itemize}

    \noindent There is only one state called \emph{loop} for Coercer.
    It has 4 looping transitions.
    Their update expressions take the form of a Boolean value assignment.
    A more common approach would be to clone the state
    for possible Boolean evaluations and find a reachable subset there.
    However, this would lead to a loss of generality and readability
    of the module (e.g., the coercer module would have $2^{3\cdot2}$ states for 3 voters, etc.).

    \smallskip\noindent Transitions with respect to Voter \texttt{v\_id} (clockwise starting with top left):
    \begin{itemize}
    	\item establish a contact with Voter, 
    	set \texttt{coercion[v\_id]} to $true$;
    	
    	\item if have not seen proof of vote, enable transition;
    	if taken, set \texttt{seen[v\_id]} to $true$;
    	
    	\item if Voter was coerced, enable transition;
    	if taken, then punish a Voter,
    	set \texttt{coercion[v\_id]} to $false$, finalizing interaction;
    	
    	\item if Voter was coerced, enable transition;
    	if taken, then do not punish a Voter,
    	set \texttt{coercion[v\_id]} to $false$, finalizing interaction.
    \end{itemize}
}

\begin{figure}[t]	
	\centering
	\includegraphics[width=\textwidth]{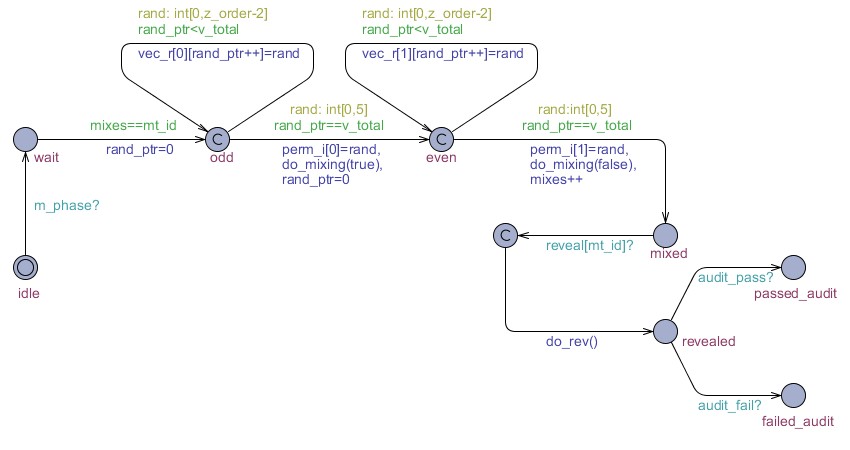}
	\caption{Mteller template}\label{fig:mteller}		
\end{figure}

\subsection{Mix Teller (Mteller)}\label{sec:mteller}

Once the mixing phase starts, each mix teller performs two re-encryption mixes.
The order of turns is ascending and determined by their identifiers.
The randomization factors and permutation of each mix are selected in a nondeterministic way and stored for a possible audit of re-encryption mixes.
When audited, the mix teller reveals the requested links and the associated factors, thus allowing Auditor to verify that the input ciphertext maps to the output.
The structure of the mix teller is shown in Figure~\ref{fig:mteller}.

\extended{
  \smallskip\noindent
  Private variables:
  \begin{itemize}
  	\item \texttt{vec\_r}: 2-dimensional integer list (size $2\times|v\_total|$)
  	of randomization factors used for re-encryption;
  	
  	\item \texttt{perm\_i}: 2-dimensional integer list (size $2\times|v\_total|$) of permutation indices used for re-encryption;
  	
  	\item \texttt{rand\_ptr}: (meta variable) index for \texttt{vec\_r};
  	
  	\item \texttt{mycol}: a pair of \texttt{board} column indices reserved
  	for a given mix teller.
  \end{itemize}

  \smallskip\noindent Procedures:
  \begin{itemize}
  	\item \texttt{do\_mixing(mi)}:
  	using board column \texttt{(mycol[mi]-1)} as an input,
  	re-encrypt each term using randomization factors from \texttt{vec\_r[mi]},
  	shuffle them with \texttt{perm\_i[mi]} permutation and paste result to \texttt{mycol[mi]};
  	
  	\item \texttt{do\_rev()}: uses shared variables $rev\_r$ and $rev\_p$
  	to reveal (pass to Auditor) randomization factors and links for the audited terms.
  	
  \end{itemize}

  \smallskip\noindent States:
  \begin{itemize}
  \item \emph{idle}: waiting for the start of the mixing phase;
  \item \emph{wait}: waiting for a turn for mixing;
  \item \emph{odd}: performing odd mix;
  \item \emph{even}: performing even mixing;
  \item \emph{mixed}: finished mixing, passed turn to the next mix teller (if any),
  waiting for a possible audit;
  \item \emph{revealed}: revealed values needed for audit,
  waiting for Auditor's verdict;
  \item \emph{passed\_audit}: mix teller passed Auditor's correctness check;
  \item \emph{failed\_audit}: mix teller failed Auditor's correctness check.
  \end{itemize}

  \smallskip\noindent Transitions:

  \begin{itemize}

  \item \emph{idle$\rightarrow$wait}: enabled transition;

  \item \emph{wait$\rightarrow$odd}: if it is current mix teller's turn,
  enable transition; if taken, then initialize \texttt{rand\_ptr} to zero;

  \item \emph{odd$\rightarrow$odd}: if new randomization factors not selected yet,
  enable transition; if taken, insert random value $rand$ into
  \texttt{vec\_r[0][rand\_ptr]} and then increment \texttt{rand\_ptr};

  \item \emph{odd$\rightarrow$even}: if randomization factors are ready,
  enable transition; randomly select permutation index, perform re-encryption mix
  using those and reset \texttt{rand\_ptr} counter to generate a new randomness;

  \item \emph{even$\rightarrow$even}: if new randomization factors not selected yet,
  enable transition; if taken, insert random value $rand$ into
  \texttt{vec\_r[1][rand\_ptr]} and then \texttt{rand\_ptr};

  \item \emph{even$\rightarrow$mixed}: if randomization factors are ready,
  enable transition; randomly select permutation index, perform re-encryption mix
  using those and pass turn incrementing \texttt{mixes};

  \item \emph{mixed$\rightarrow$audit}: enabled transition;
  if taken, then mix teller will be in committed state,
  from where will have to reveal mix factors for audited terms;

  \item \emph{revealed$\rightarrow$(passed\_audit {\normalfont ||} failed\_audit)}: enabled transition.
  \end{itemize}
}

\begin{figure}[t]	
	\centering
	\includegraphics[width=\textwidth]{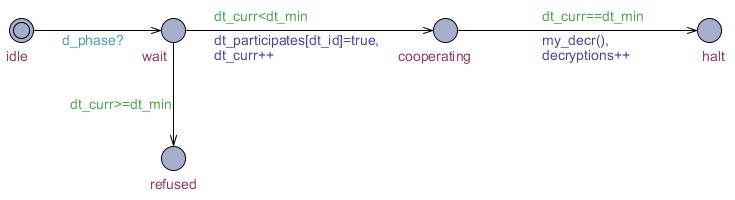}
	\caption{Dteller template}\label{fig:dteller}		
\end{figure}

\subsection{Decryption Teller (Dteller)}\label{sec:dteller}

In this module, after the re-encryption mixes are done, a decryption teller nondeterministically chooses a subset of tellers.
Note that it is possible that two or more of decryption tellers refuse to cooperate, which would lead to a deadlock.
In order to avoid that, only subsets with cardinality of $2$ will be considered.
%

\extended{
  We will use \textit{Shamir (2,3)-Threshold Scheme} for decryption.
  Consider a polynomial $a(x)$ of a degree $1$,
  such that $a(0)=k$, where $k$ is a secret key.
  Each decryption teller $d_i\in\{0,1,2\}$ will have a point $(x_{d_i},y_{d_i})$ on that polynomial,
  where $x_{d_i}$ is publicly known and $y_{d_i}=a(x_{d_i})$ is a key share.
  In order to reconstruct the secret $k$,
  a group of $2$ tellers will have to cooperate,
  using Lagrange interpolation formula.
  We assume that a polynomial $a(x)$ was set and
  secret shares were assigned to each participant in advance.\smallskip
}

\extended{
  \smallskip\noindent Private variables:
  \begin{itemize}
  \item \texttt{k\_share}: the teller's share of the secret;
  \item \texttt{x}: the value of the first variable in pair $(x,y)$.
  \end{itemize}

  \smallskip\noindent Procedures:
  \begin{itemize}
  \item \texttt{my\_decr()}: depending on the set of participants,
  multiply $k\_share$ by a proper Lagrange basis and use the result
  as a key for decryption of an input column.
  To determine the set of participants,
  a shared Boolean list \texttt{$dt\_paricipants$} is used.
  \end{itemize}

  \smallskip\noindent States:
  \begin{itemize}
  \item \emph{idle}: waiting for the decryption phase;
  \item \emph{wait}: wait for one's turn to make a decision;
  \item \emph{refused}: refused to cooperate;
  \item \emph{cooperating}: will participate in decryption;
  \item \emph{halt}: finished his decryption.
  \end{itemize}

  \smallskip\noindent Transitions:
  \begin{itemize}
  \item \emph{idle$\rightarrow$wait}: enabled transition;

  \item \emph{wait$\rightarrow$refused}: if number of participants is already enough
   for key reconstruction, enable transition;

  \item \emph{wait$\rightarrow$cooperating}: if the number of participants is less than
  required for key reconstruction, enable transition;
  if taken, then set \texttt{dt\_paricipants[dt\_id]} to $true$ and increment
  the current number of participants \texttt{dt\_curr};

  \item \emph{cooperating$\rightarrow$halt}: if the number of participants is sufficient,
  enable transition; if taken, then proceed to decryption.
  \end{itemize}
}

\begin{figure}[t]	
	\centering
	\includegraphics[width=\textwidth]{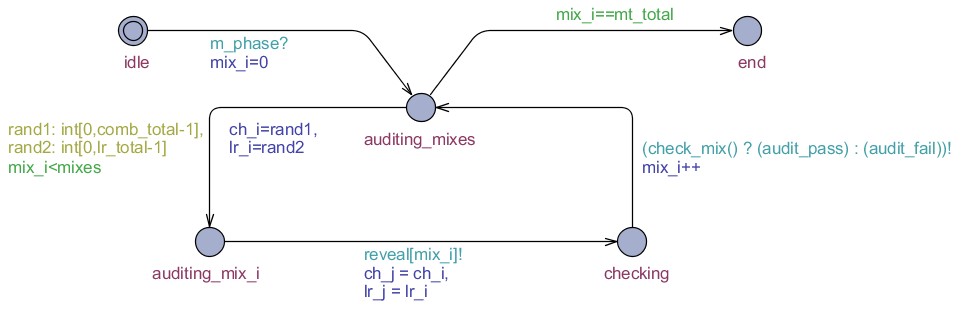}
	\caption{Auditor template}\label{fig:auditor}		
\end{figure}

\subsection{Auditor}\label{auditor}

In order to confirm that the mix tellers performed their actions correctly,
the Auditor conducts an audit based on the randomized partial checking technique, RPC in short~\cite{Jakobsson02mixnets}.
To this end, he requests each mix teller to reveal the factors for the selected half of an odd-mix batch, and verify whether the input corresponds to the output.
\extended{
  The choice to audit in-phase or after the mixing is nondeterministic.
  Possible selections for terms and sides of links are encoded as elements of a list.

}
The control flow of the Auditor is presented in Figure~\ref{fig:auditor}.
\extended{

  \smallskip\noindent Private variables:
  \begin{itemize}
  \item \texttt{mix\_i}: index of currently audited mix teller;

  \item \texttt{ch\_i}: index of \texttt{audit\_ch} terms combination (a subset for audit);

  \item \texttt{lr\_i}: index of \texttt{audit\_lr} splitting, used for the left and right linkage reveal.
  \end{itemize}

  \smallskip\noindent Procedures:
  \begin{itemize}
  \item \texttt{check\_mix()}: return true if audited terms correspond to
  encryption of linked ones from \texttt{rev\_p} using randomization factors
  from \texttt{rev\_r}, otherwise return false.
  \end{itemize}

  \smallskip\noindent States:
  \begin{itemize}
  \item \textit{idle}: wait for the start of mixing phase;

  \item \textit{auditing\_mixes}: in a process of auditing mix tellers;

  \item \textit{auditing\_mix\_i}: in a process of auditing mix teller $mix\_i$;

  \item \textit{cheking}: received revealed link and randomization factors from
  mix teller, can now proceed to correctness check;

  \item \textit{end}: finished mix tellers audit.
  \end{itemize}

  \smallskip\noindent Transitions:
  \begin{itemize}
  \item \textit{idle$\rightarrow$auditing\_mixes}: enabled transition;
  if taken, then set \texttt{mix\_i} counter to $0$;

  \item \textit{auditing\_mixes$\rightarrow$auditing\_mix\_i}:
  if have not audited all mix tellers, enable transition;
  if taken, then randomly select indices \texttt{ch\_i} for batch subset
  and \texttt{lr\_i} for left-right split;

  \item \textit{auditing\_mix\_i$\rightarrow$checking}:
  pass the \texttt{ch\_i} and \texttt{lr\_i} indices to Mix Teller \texttt{mix\_i}
  over channel \texttt{reveal[mix\_i]}
  using shared variables \texttt{ch\_j} and \texttt{lr\_j};

  \item \textit{checking$\rightarrow$auditing\_mixes}:
  depending on result of \texttt{check\_mix()} procedure,
  pass the correctness check verdict to Mix teller using
  either \texttt{audit\_pass} or \texttt{audit\_fail},
  then increment \texttt{mix\_i} counter by 1;

  \item \textit{auditing\_mixes$\rightarrow$end}:
  if all Mix tellers were audited, enable transition.

  \end{itemize}
}

\begin{figure}[t]	
	\centering
	\includegraphics[width=\textwidth]{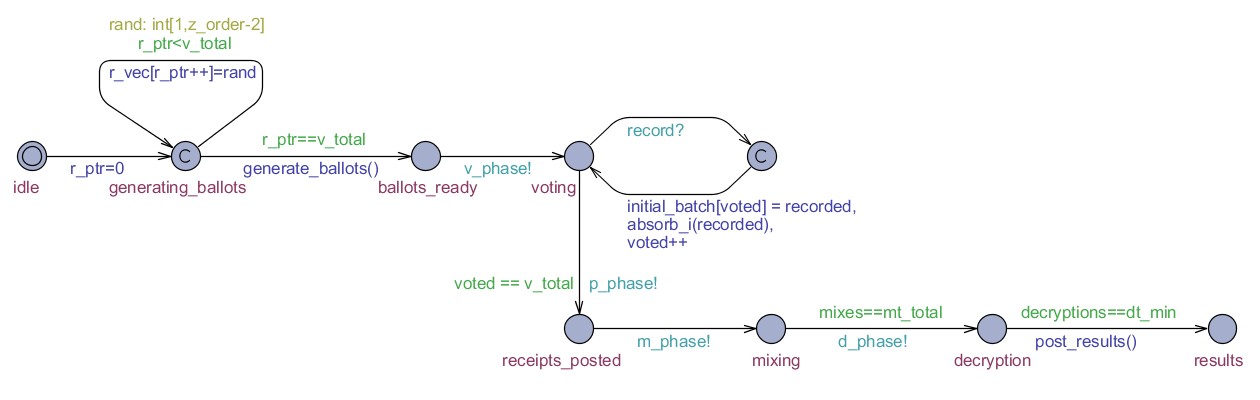}
	\caption{Sys Module}\label{fig:sys}		
\end{figure}

\subsection{Voting Infrastructure Module (Sys)}\label{sys}

This module represents the behavior of the election authority that prepares the ballot forms, monitors the current phase, signals the progress of the voting procedure to the other components, and posts the results of the election at the end.
In addition, the module plays the role of a server that receives receipts and transfers them to the database throughout the election.

\extended{
  \smallskip\noindent Private variables:
  \begin{itemize}
  \item \texttt{vote\_sum}: list of integers,
  that maps candidates to the sum of votes they have;

  \item  \texttt{r\_vec}: list of randomization factors, used for
  encrypting the seed into onion during ballot generation;

  \item \texttt{r\_ptr}: (meta variable) index of \texttt{r\_vec} element;

  \item \texttt{voted}: counter of scanned receipts.
  \end{itemize}

  \smallskip\noindent Procedures:
  \begin{itemize}
  \item \texttt{generate\_ballots()}: encrypt the list of seeds using
  randomization factors from \texttt{r\_vec};

  \item  \texttt{absorb\_i(recorded)}: absorb the index $i$ of the marked cell
  in the \texttt{recorded} receipt into its onion and paste the result to the board;

  \item \texttt{post\_results() }: map terms of the form $g^{r+s}$ from the last column to the candidate $x$ using formula $x=(r+s)\ (mod\ c\_total)$
  and a look-up table \texttt{dlog}.
  \end{itemize}

  \smallskip\noindent States:
  \begin{itemize}
  \item \textit{idle}: initial state;

  \item \textit{generating\_ballots}: there are either no ballots or they are
  being generated at the moment;

  \item \textit{ballots\_ready}: all the ballots have been  prepared;

  \item \textit{voting}: election phase, when voters can obtain ballot forms and cast their votes;

  \item \textit{receipts\_posted}: the batch of initial receipts
  is now publicly seen and can be checked by the voters;

  \item \textit{mixing}: wait for the re-encryption mixes to finish;

  \item \textit{decryption}: wait for the decryption to finish;

  \item \textit{results}: the tally is posted.
  \end{itemize}

  \smallskip\noindent Transitions:
  \begin{itemize}
  \item \textit{idle$\rightarrow$generating\_ballots}: enabled transition;
  if taken, then reset \texttt{r\_ptr} counter
  to generate a list of randomization factors;

  \item \textit{generating\_ballots$\rightarrow$generating\_ballots}:
  if randomization factors not prepared, enable transition;
  if taken, then insert a random value \texttt{rand} into
  \texttt{r\_vec[r\_ptr]} and increment \texttt{r\_ptr} iterator;

  \item \textit{generating\_ballots$\rightarrow$ballots\_ready}:
  if randomization factors are prepared, enable transition;
  if taken, generate ballots using randomization factors from \texttt{r\_vec};

  \item \textit{ballots\_ready$\rightarrow$voting}: broadcast the start of
  voting phase to Voters using \texttt{v\_phase} channel;

  \item \textit{voting$\rightarrow \rightarrow$voting}: receive a receipt from the voter
  and from the committed state append it to the board,
  then increment a counter of receipts scanned;

  \item \textit{voting$\rightarrow$receipts\_posted}:
  if all voters submitted receipts, enable transition;
  if taken, then broadcast that Voters that initial receipts were posted using the \texttt{p\_phase} channel;

  \item \textit{receipts\_posted$\rightarrow$mixing}:
  broadcast the start of mixing phase to mix tellers using \texttt{m\_phase} channel;

  \item \textit{mixing$\rightarrow$decryption}: if all mixes are
  done, enable transition;
  if taken, then broadcast the start of decryption phase to decryption tellers
  using \texttt{d\_phase} channel;

  \item \textit{decryption$\rightarrow$results}: if all decryptions were done, enable
  transition; if taken, then calculate and post results tally.
  \end{itemize}
}

%
%
%



%% file: verification.tex
As we already mentioned, we chose \Uppaal for this study mainly because of its modeling functionality.
The requirement specification capabilities of the tool were of secondary importance.
In fact, they are quite limited.
First, \Uppaal admits only a fragment of \CTL: it excludes the ``next'' and ``until'' modalities, and does not allow for nesting of operators (with one exception that we describe below).
Thus, the supported properties fall into the following categories:\
simple \emph{reachability} ($\Epath\Sometm\prop{p}$),
\emph{liveness} ($\Apath\Sometm\prop{p}$),
and \emph{safety} ($\Apath\Always\prop{p}$ and $\Epath\Always\prop{p}$).
The only allowed nested formulas come in the form of the \emph{$p$ leads to $q$}
property, written $\prop{p}\leadsto\prop{q}$, and being a shorthand for $\Apath\Always(\prop{p}\then\Apath\Sometm\prop{q})$.

Still, \Uppaal allows to model-check simple properties of \Pret, as we show in Section~\ref{sec:mcheck-basic}.
Moreover, by tweaking models and formulas in a clever way, one can also verify some more sophisticated requirements, see Section~\ref{sec:mcheck-CTLK}.

\subsection{Model Checking Temporal Requirements}\label{sec:mcheck-basic}

The model in Section~\ref{sec:pret-model} allows us to verify properties of \Pret for different configurations of participants. In particular, we have analyzed variants of the model, generated by different numbers of instances for the Voter template.
The values should be chosen carefully, to avoid state-space explosion.
We have considered the following properties for verification:
\begin{enumerate}
 \item\label{it:mixteller} $\Epath\Sometm\prop{failed\_audit_0}$: the first mix teller might eventually fail an audit.
 \item\label{it:punished} $\Apath\Always\neg\prop{punished_i}$: voter $i$ will never be punished by the coercer.
 \item\label{it:markedchoice} $\prop{has\_ballot_i\leadsto marked\_choice_i}$: on all paths, whenever voter $i$ gets a ballot form, she will eventually mark her choice.
\end{enumerate}

We verified each formula on several configurations differing by the number of voters ranging from 1 to 5.
For the first property, the \Uppaal verifier returns `Property is satisfied' for the configurations with $1,2,3$ and $4$ voters. In case of $5$ voters,
we get `Out of memory.'
Formula~(\ref{it:punished}) produces the answer `Property is not satisfied' and pastes a counter-example into the simulator (if the diagnostic trace was set in the option panel) for all the five configurations.
Finally, formula~(\ref{it:markedchoice}) ends with `Out of memory' regardless of the number of voters.

\extended{
  \para{Optimizations.} 
  To keep the model manageable and in attempt to reduce the state space, every numerical variable is defined as a bounded integer in a form
  of \texttt{int[min,max]}, restricting its range of values.\footnote{
    Without the explicit bounds, the range of values would be [-32768,32768]. }
  The states violating the bounds are discarded at run-time.
  For example, transition \emph{has\_ballot$\rightarrow$marked\_choice} of the Voter (Figure~\ref{fig:voter})
  has a selection of value \texttt{X} in the assignment of variable \texttt{chosen}.
  The type of \texttt{X} is \texttt{c\_t}, which is an alias to \texttt{int[0,c\_total-1]}, i.e., the range of meaningful candidate choices.

  We also tried to keep the number of used variables minimal, as it plays an important role in model checking procedure.
}


\begin{figure}[t]
\begin{center}
\begin{tabular}{c@{\quad}c@{\qquad\qquad\qquad}c@{\quad}c}
  \begin{tabular}{c}
  (a)
  \end{tabular}
&
  \begin{tabular}{c}
	\input{images/triangle_ep.tex}
  \end{tabular}
&
  \begin{tabular}{c}
  (b)
  \end{tabular}
&
  \begin{tabular}{c}
	\input{images/triangle_rev.tex}
  \end{tabular}
\end{tabular}
\end{center}
\caption{(a) Epistemic bisimulation triangle;\ (b) turning the triangle into a cycle by reversing the transition relation}\label{fig:triangle}
\end{figure}

\begin{figure}[t]	
	\hspace{-1cm}
	\includegraphics[width=1.15\textwidth]{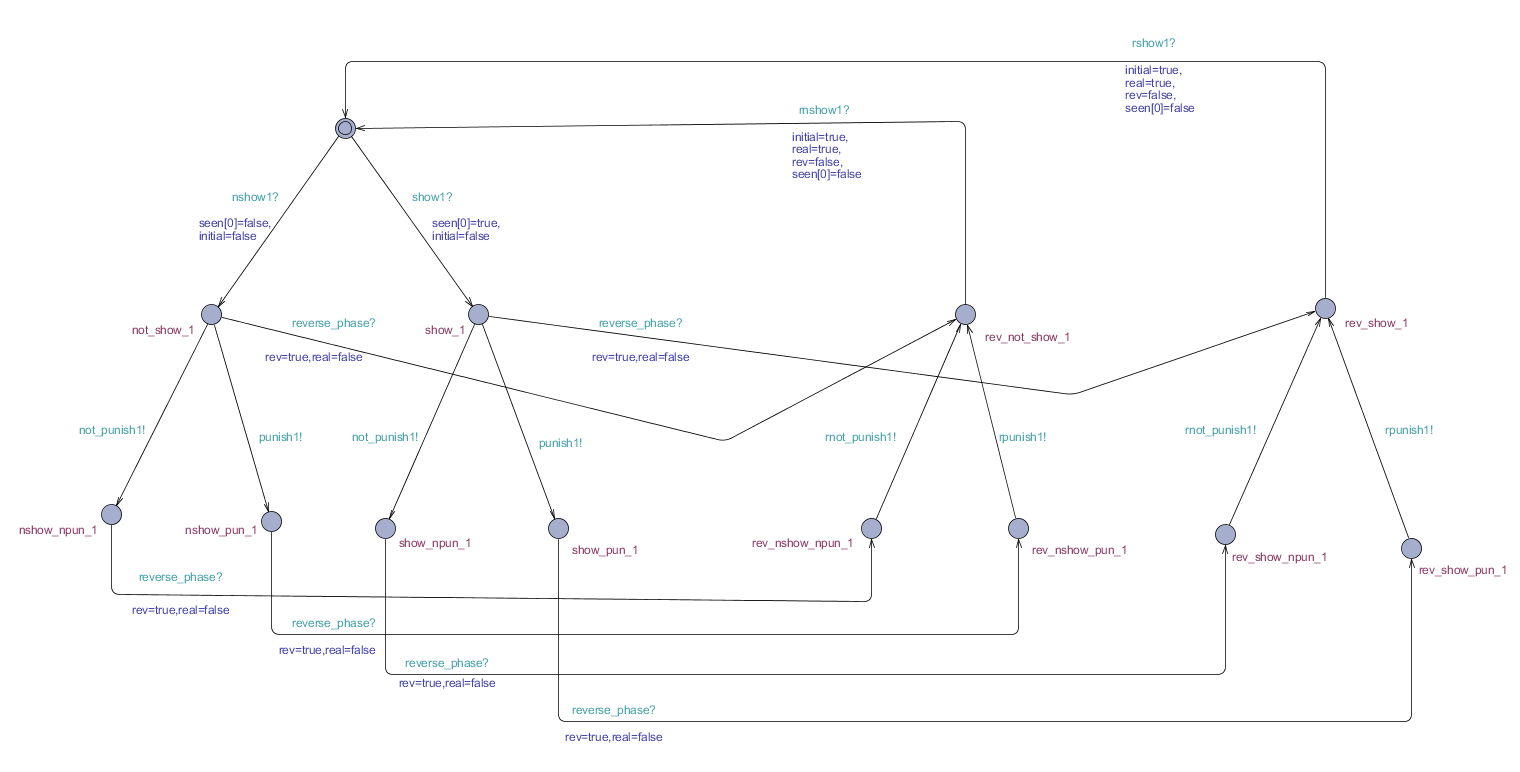}
	\caption{Coercer module augmented with the converse transition relation}\label{fig:rev_coercer}
\end{figure}

\extended{
  \begin{figure}[t]	
  	\centering
  	\includegraphics[width=\textwidth]{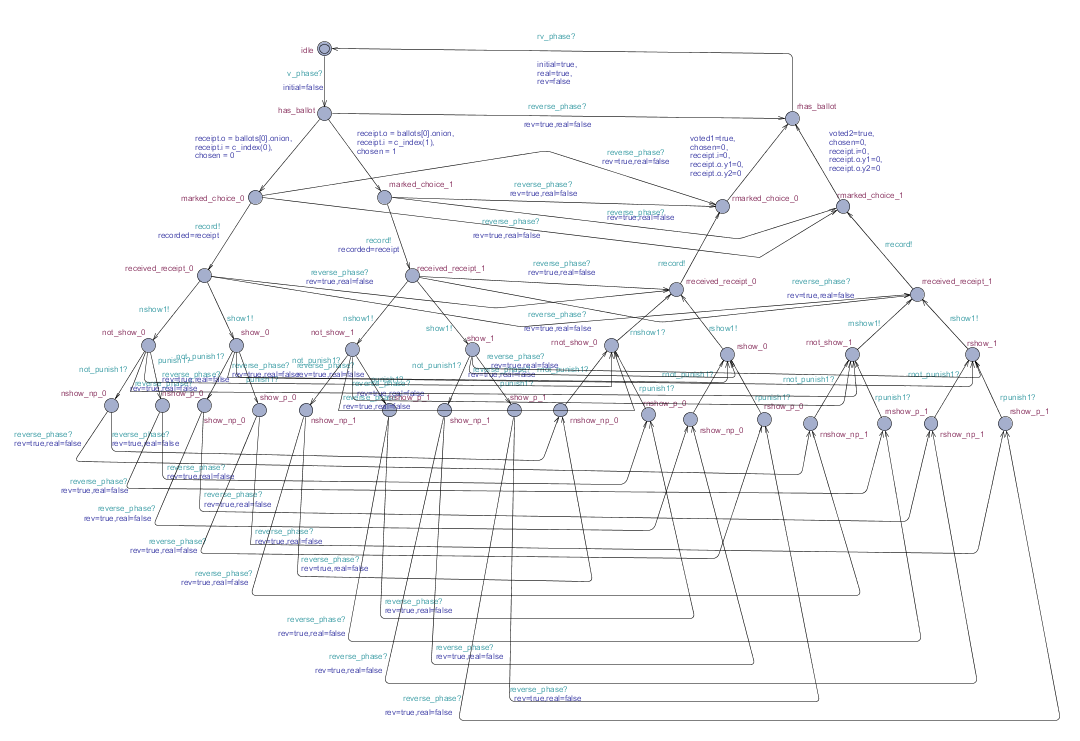}
  	\caption{Voter1 with reversed transitions}\label{fig:rev_voter1}		
  \end{figure}
}

\subsection{How to Make Model Checker Do More Than It Is Supposed To}\label{sec:mcheck-CTLK}

Many important properties of voting refer to the knowledge of its participants. For example, receipt-freeness expresses that the coercer should never know how the voter has voted. Or, better still, that the coercer will never know if the voter voted as instructed.
Similarly, voter-verifiability says that the voter will eventually know whether her vote has been registered and tallied correctly (assuming that she follows the verification steps).

A clear disadvantage of \Uppaal is that its language for specification of requirements is restricted to purely temporal properties.
We show that, with some care, one can use it to embed the verification of more sophisticated properties. In particular, we show how to enable verification of some knowledge-related requirements by a technical reconstruction of models and formulas. The construction has been inspired by the reduction of epistemic properties to temporal properties, proposed in~\cite{Goranko04comparingKRA,Jamroga08reduction}. Consequently, \Uppaal and similar tools can be used to model check formulas of \CTLK (i.e., \CTL + Knowledge) that express variants of receipt-freeness and voter-verifiability.

In order to simulate the knowledge operator $K_a$ under the \CTL semantics, the model needs to be modified.
The first step is to understand how the formula 
$\lnot K_c \neg\prop{voted_{i,j}}$
(saying that the coercer doesn't know that the voter hasn't voted for candidate $j$)
is interpreted.
Namely, if there is a reachable state in which $\prop{voted_{i,j}}$ is true, there must also exist another reachable state, which is indistinguishable from the current one, and in which
$\neg\prop{voted_{i,j}}$ holds. The idea is shown in Figure~\ref{fig:triangle}a.
We observe that to simulate the epistemic relation we need to create copies of the states in the model (the ``real'' states). We will refer to those copies as the \emph{reverse states}. They are the same as the real states, but with reversed transition relation. Then, we add transitions from the real states to their corresponding reverse states, that simulate the epistemic relation between the states. This is shown in Figure~\ref{fig:triangle}b.

To illustrate how the reconstruction of the model works on a concrete example, we depict the augmented \short{Coercer }template\extended{s} in Figure\extended{s}~\ref{fig:rev_coercer}\extended{--\ref{fig:rev_system}}.

\extended{
  \begin{figure}[t]	
  	\centering
  	\includegraphics[width=\textwidth]{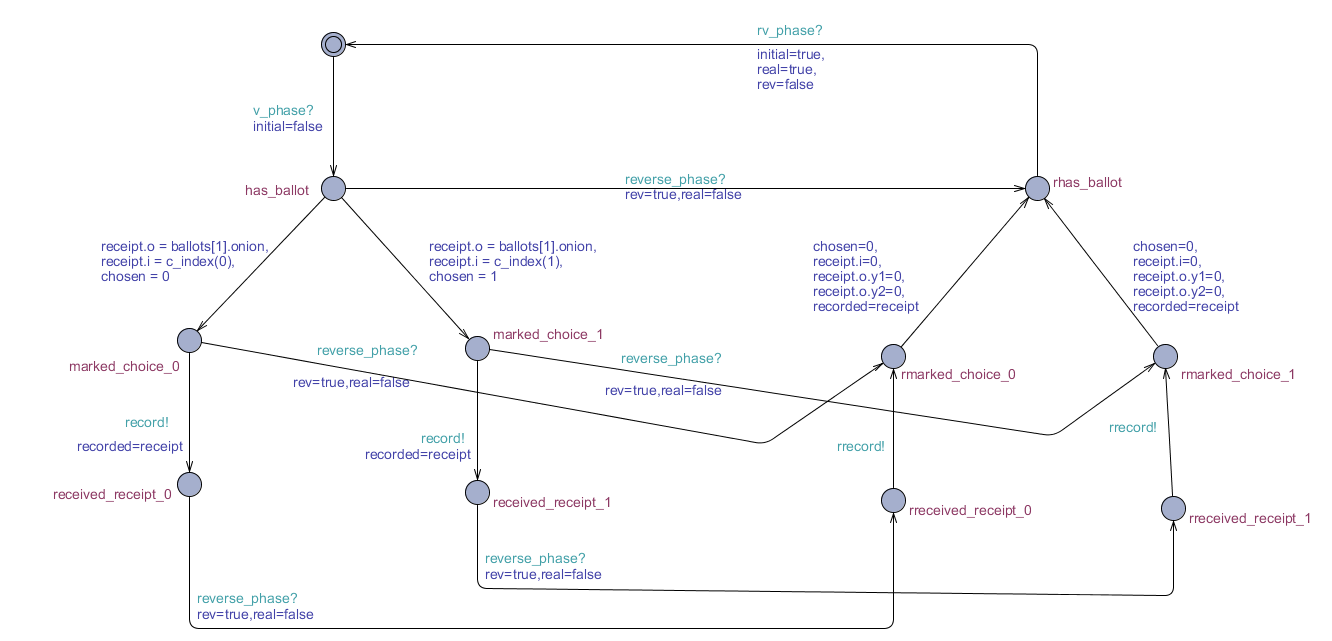}
  	\caption{Voter2 with reversed transitions}\label{fig:rev_voter2}
  \end{figure}

  \begin{figure}[t]	
  	\centering
  	\includegraphics[width=\textwidth]{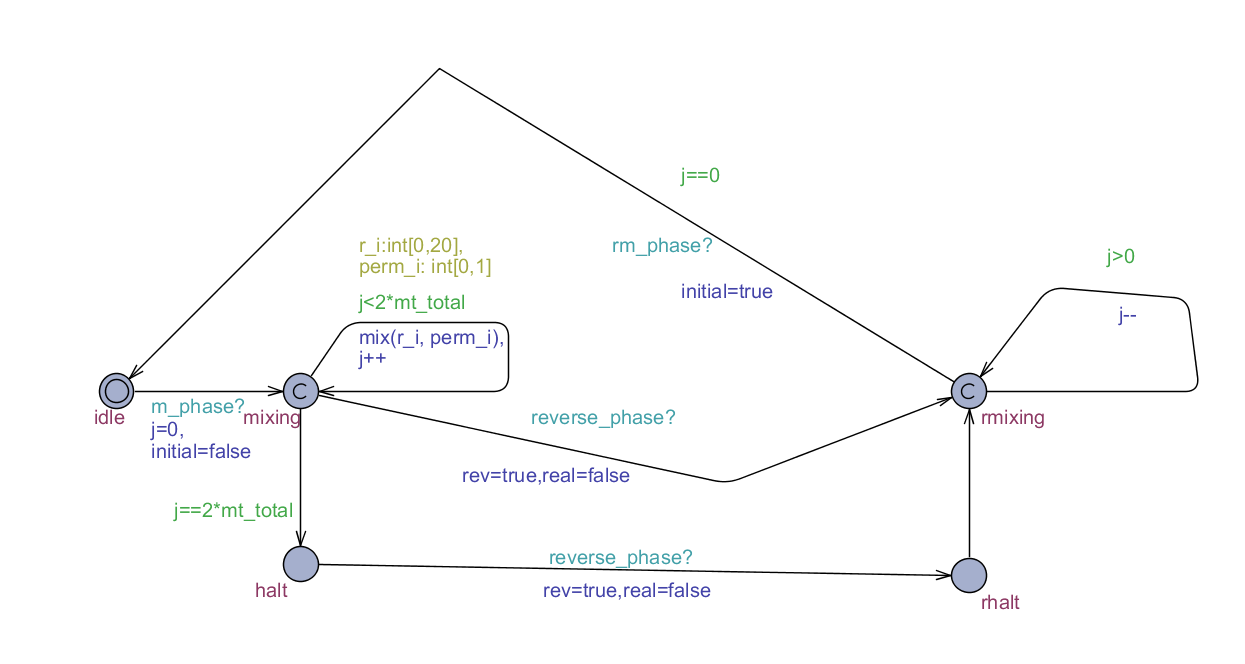}
  	\caption{Mix Teller with reversed transitions}\label{fig:rev_mteller}
  \end{figure}

  \begin{figure}[t]	
  	\centering
  	\includegraphics[width=0.7\textwidth]{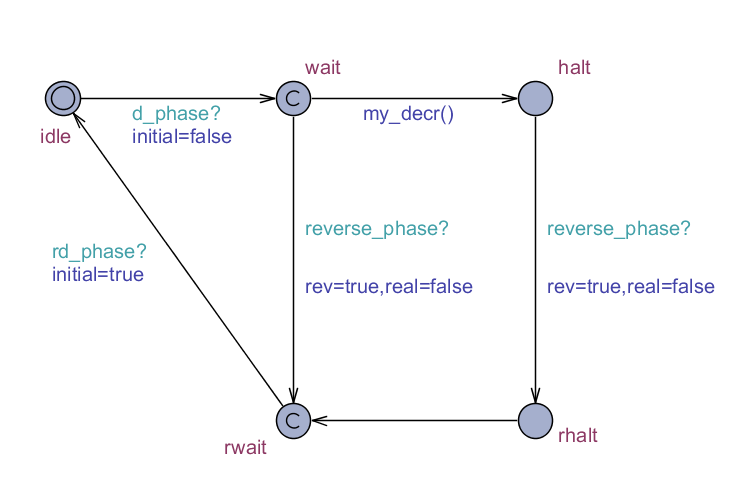}
  	\caption{Decryption Teller with reversed transitions}\label{fig:rev_dteller}		
  \end{figure}

  \begin{figure}[t]	
  	\centering
  	\includegraphics[width=\textwidth]{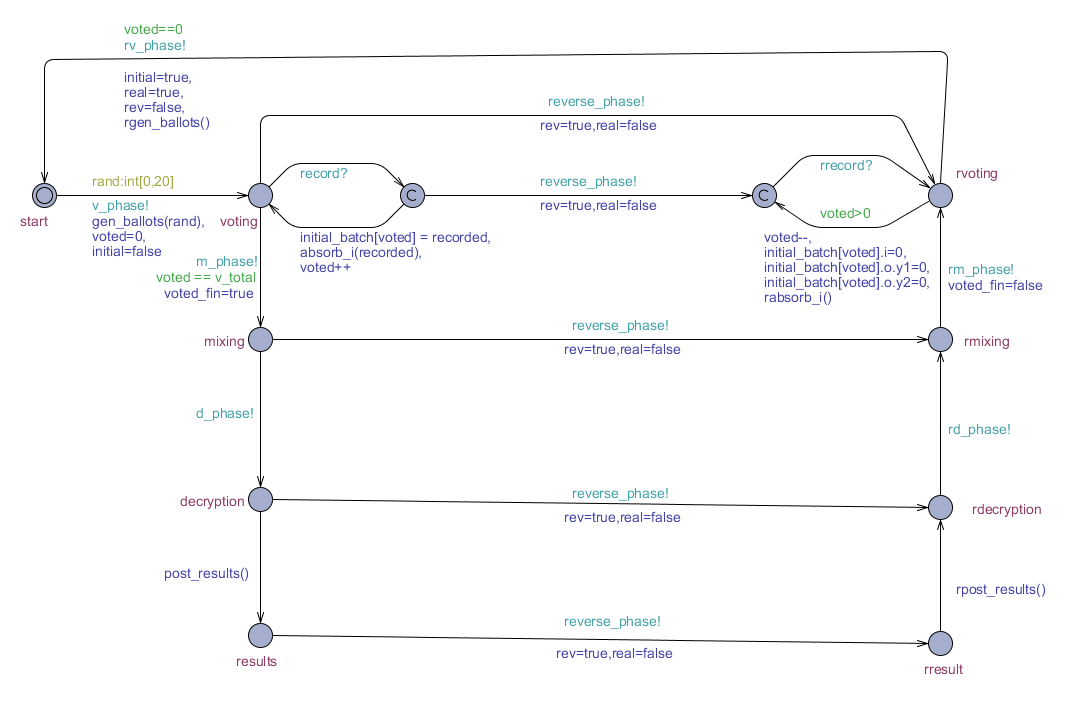}
  	\caption{Sys module with reversed transitions}\label{fig:rev_system}		
  \end{figure}
}


The next step is the reconstruction of formulas.
Let us take the formula for the weak variant of receipt-freeness from Section~\ref{sec:uppaal-formulas}, i.e., $\Epath\Sometm(\prop{results} \land \neg\prop{voted_{i,j}} \land \neg K_c \neg\prop{voted_{i,j}})$.
In order to verify the formula in \Uppaal, we need to replace the knowledge operator according to our model reconstruction method (see Figure~\ref{fig:triangle} again). This means that the verifier should find a path that closes the cycle: from the initial state, going through the real states of the voting procedure to the vote publication phase, and then back to the initial state through the reversed states. In order to ``remember'' the relevant facts along the path, we use persistent Boolean variables $\prop{voted_{i,j}}$ and $\prop{negvoted_{i,j}}$: once set to true they always remain true. We also introduce a new persistent variable $\prop{epist\_voted_{i,j}}$ to refer to the value of the vote after an epistemic transition. Once we have all that, we can propose the reconstructed formula: $\Epath\Sometm(\prop{results} \land \prop{negvoted_{i,j}} \land \prop{epist\_voted_{i,j}} \land \prop{initial})$. \Uppaal reports that the formula holds in the model.


A stronger variant of receipt-freeness is expressed by another formula from Section~\ref{sec:uppaal-formulas}, i.e., $\Apath\Always(\prop{results} \then \neg K_c \neg\prop{voted_{i,j}})$. Again, the formula needs to be rewritten to a pure \CTL formula. As before, the model checker should find a cycle from the initial state, ``scoring'' the relevant propositions on the way. More precisely, it needs to check if, for every real state in which election has ended, there exist a path going back to the initial state through a reverse state in which the voter has voted for the selected candidate. This can be captured by the following formula: $\Apath\Always\big((\prop{results} \land \prop{real}) \then \Epath\Sometm(\prop{voted_{i,j}} \land \prop{init})\big)$. Unfortunately, this formula cannot be verified in \Uppaal, as \Uppaal does not allow for nested path quantifiers. In the future, we plan to run the verification of this formula using another model checker LTSmin~\cite{Kant15ltsmin} that accepts \Uppaal models as input, but allows for more expressive requirement specifications.

%% file: images/triangle_ep.tex

		
		\begin{tikzpicture}[transform shape, scale = 0.60, decoration={snake, pre length=3pt, post length=7pt}]
    \node[state, initial=left] (s0) at (0.0, 0.0) {$init$};
    \node[state, fill=green!20] (s1) at (0.0, -4.0) {$voted1$};
    \node[state, fill=red!20] (s2) at (4.0, -4.0) {$voted2$};
		
		\path[->, draw=green, very thick, decorate] (s0) -- node[left] {} (s1);
		\path[->, draw=green, very thick, decorate] (s0) -- node[right] {} (s2);
    
    \path [-,style=dashed,shorten >=1pt, auto, node distance=7cm, semithick]
    (s1) edge node {Coercer} (s2)
    ;
  \end{tikzpicture}

%% file: images/triangle_rev.tex

		
		\begin{tikzpicture}[transform shape, scale = 0.60, decoration={snake, pre length=3pt, post length=7pt}]
    \node[state, initial=left] (s0) at (0.0, 0.0) {$init$};
    \node[state, fill=green!20] (s1) at (0.0, -4.0) {$voted1$};
    \node[state, fill=red!20] (s2) at (4.0, -4.0) {$r\_voted2$};
		
		\path[->, draw=green, very thick, decorate] (s0) -- node[left] {real} (s1);
		\path[->, draw=red, very thick, decorate] (s2) -- node[right] {reverse} (s0);
    
    \path [->,style=solid,shorten >=1pt, auto, node distance=7cm, very thick]
    (s1) edge node {epistemic} (s2)
    ;
  \end{tikzpicture}

%% file: extensions.tex
\begin{figure}[t]	
	\centering
	\includegraphics[width=\textwidth]{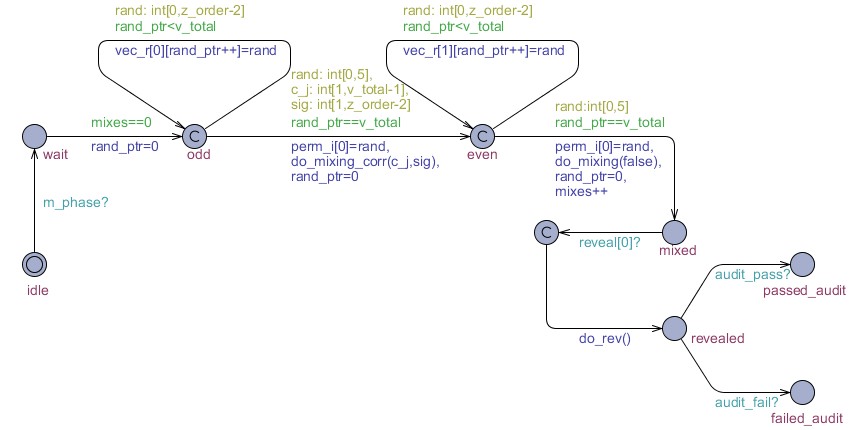}
	\caption{Corrupted Mix Teller module}\label{fig:corr_mteller}		
\end{figure}

A version of \emph{Pfitzmann's attack} is known to compromise mix-nets with randomized partial checking~\cite{Khazaei13rpc}.
It can be used to break the privacy of a given vote with probability $1/2$ of being undetected during RPC.
The leaked information may differ depending on both the implementation of the attack and the voting protocol.

The main idea is that the first mix teller, who is corrupted, targets a ciphertext $c_i$ from the odd mix input,
and replaces some output term $c_j$ with $c_i^{\delta}$, where $\delta$ is a properly chosen integer (the criteria depend on the implementation).
After the decryption results are posted, a pair of decrypted messages $m$ and $m'$ satisfying equation $m'=m^{\delta}$ can be used to deduce sensitive information.

Clearly, the model presented in Section~\ref{sec:pret-model} is too basic to allow for detection of the attack.
Still, this can be obtained by a simple extension of the model. For that, we change the Mteller template as shown in Figure~\ref{fig:corr_mteller}.
The only difference lies in how the first re-encryption mix is done: the corrupted mix teller targets $c_0$, chooses a random non-zero $\delta$, and uses $c_0^{\delta}$ instead of some other output term. In all other respects, the teller behaves honestly.

Using the \Uppaal model checking functionality, it can be verified that there exist executions where the corrupted mix teller has failed the audit, as well as paths where he has passed. That is, both $\Epath\Sometm\prop{failed\_audit_0}$ and $\Epath\Sometm\prop{passed\_audit_0}$ produce `Property satisfied' as the output.
We note that, in order to successfully verify those properties in our model of \Pret, the search order option in \Uppaal should be changed from the (default) Breadth First to either Depth First or Random Depth First.

%% file: relatedwork.tex

\extended{\para{Coercion-resistant and voter-verifiable voting systems.}}
Over the years, the properties of \emph{ballot secrecy}, \emph{receipt-freeness}, \emph{coercion resistance}, and \emph{voter-verifiability} were recognized as important for an election to work properly\extended{.
In particular, {receipt-freeness} and {coercion-resistance} were studied and formalized in multiple ways~\cite{Benaloh94receipt,Okamoto98receipt,Delaune06coercion,Kusters10game,Dreier12formal}}, see\extended{ also}~\cite{Meng09critical\extended{,Tabatabaei16expressing}} for an overview.
\extended{
  More recently, significant progress has been made in the development of voting systems that would be coercion-resistant and at the same time allow the voter to verify ``her'' part of the election outcome~\cite{Ryan15verifiability,Cortier16sok-verifiability}.
}
A number of secure and voter-verifiable schemes have been proposed, most notably the \Pret protocol for supervised elections~\cite{\extended{Chaum05pretavoter,}Ryan10atemyvote}, the Pretty Good Democracy approach to internet voting~\cite{Ryan13prettygood}, and Selene, an enhanced form of tracking number-based scheme~\cite{Ryan16selene}.
\extended{

}%
Nowadays, such schemes are starting to move out of the laboratory and into use in real elections.  For example, \Pret has been successfully used in one of the state elections in Australia~\cite{Burton12elections-victoria} while the Scantegrity II system~\cite{Chaum09scantegrityII} was used in municipal elections in the Takoma Park county, Maryland.
\extended{
  Moreover, a number of verifiable schemes were used in non-political elections. E.g., Helios~\cite{Adida08helios} was used to elect officials of the International Association of Cryptologic Research and the Dean of the University of Louvain la Neuve.
}
This strongly emphasizes the need for extensive analysis and validation of such systems.

\extended{
  \para{Formal verification of voting protocols.}
  In voting systems, \emph{verifiable} means that the voters are able to verify the outcome of the election. This is different from \emph{formal verification} whose task is to provide algorithmic tools for checking if a system satisfies a given requirement. General verification tools for security protocols include ProVerif~\cite{Cheval13proverif}, Scyther~\cite{Cremers12scyther}, AVISPA~\cite{Armando05avispa-short}, and Tamarin~\cite{Meier13tamarin}.

}
Formal analysis of selected voting protocols, based on theorem proving in first-order logic or linear logic, includes attempts at verification of vote counting in~\cite{Beckert13votecounting,Pattinson15votecounting}.
The Coq theorem prover\extended{ for higher-order logic}~\cite{Bertot04coq} was used to implement the STV counting scheme in a provably correct way~\cite{Ghale18stv-coq}, and to produce a provably voter-verifiable variant of the Helios protocol~\cite{Haines19verified-verifiers}.
Moreover, Tamarin~\cite{Meier13tamarin} was used to verify receipt-freeness in Selene~\cite{Bruni17selene} and Electryo~\cite{Zollinger17electryo}.
Approaches based on model checking are fewer and include the analysis of risk-limiting audits~\cite{Beckert16audits} with the CBMC model checker~\cite{Clarke04cbmc}.%
\short{
  Moreover,~\cite{Jamroga18Selene} proposed and verified a simple multi-agent model of Selene using \mcmas~\cite{Lomuscio17mcmas}.
}%
\extended{

  \para{Multi-agent models and model checkers in verification of security.}
  Multi-agent model checking is virtually unexplored in analysis of voting systems. The only relevant work that we are aware of is~\cite{Jamroga18Selene} where a simple multi-agent model of Selene was proposed and verified using the \mcmas model checker~\cite{Lomuscio17mcmas}.
}
Related research includes the use of multi-agent methodologies to specify and verify properties of authentication and key-establishment protocols~\cite{Lomuscio08securityprots,Boureanu16verifSecurity} with \mcmas.
\extended{
  In particular, \cite{Boureanu16verifSecurity} used \mcmas to obtain and verify models, automatically synthesized from high-level protocol description languages such as CAPSL, thus creating a bridge between multi-agent and process-based methods.
}

In all the above cases, the focus is on the verification itself. Indeed, all the tools mentioned above provide only a text-based interface for specification of the system. As a result, their model specifications closely resemble programming code, and insufficiently protect from the usual pitfalls of programming: unreadability of the code, lack of modularity, and opaque control structure.
In this paper, we draw attention to tools that promote modular design of the model, emphasize its control structure, and facilitate inspection and  validation.

%% file: conclusions.tex

Formal methods are well established in proving (and disproving) the correctness of cryptographic protocols.
What makes voting protocols special is that they prominently feature human and social aspects.
In consequence, an accurate specification of the behaviors admitted by the protocol is far from straightforward.
An environment that supports the creation of modular, compact, and -- most of all -- readable specifications can be an invaluable help\extended{ in the design and validation of voting systems}.

In this context, the \Uppaal model checker has a number of advantages. Its modeling language encourages modular specification of the system behavior. It provides flexible data structures, and allows for parameterized specification of states and transitions.
Last but not least, it has a user-friendly GUI. Clearly, a good graphical model helps to {understand} how the voting procedure works, and allows for a preliminary validation of the system specification just by looking at the graphs. Anybody who ever inspected a text-based system specification or the programming code itself will know what we mean.

In this paper, we try to demonstrate the advantages of \Uppaal through a case study based on the \Pret protocol.
The models that we have obtained are neat, easy to read, and easy to modify.
On the other hand, \Uppaal has not performed very well with the verification itself. This was largely due to the fact that its requirement specification language turned out to be very limited -- much more than it seemed at the first glance. We managed to partly overcome the limitations by a smart reconstruction of models and formulas. In the long run, however, a more promising path is to extend the implementation of verification algorithms in \Uppaal so that they handle nested path quantifiers and knowledge modalities, given explicitly in the formula.